\documentstyle[eqsecnum,floats,prd,aps,twocolumn]{revtex}

\newcommand{\be}{\begin{equation}}
\newcommand{\ba}{\begin{eqnarray}}
\newcommand{\ee}{\end{equation}}
\newcommand{\ea}{\end{eqnarray}}

\begin{document}
\title{Chaos in the Einstein-Yang-Mills Equations}
\author{John D. Barrow$^1$ and Janna Levin$^2$}
\address{$^{\ ^1}$Astronomy Centre, University of Sussex, 
Brighton BN1 9QH, U.K}
\address{$^2$Center for Particle Astrophysics, Le Conte Hall
UC Berkeley, Berkeley, CA 94720-7304, USA.}
\twocolumn[
\maketitle
\widetext
\date{}
\begin{abstract}
Yang-Mills color fields evolve chaotically in an anisotropically expanding
universe. The chaotic behaviour differs from that found in anisotropic
Mixmaster universes. The universe isotropizes at late times, approaching the
mean expansion rate of a radiation-dominated universe. However, small
chaotic oscillations of the shear and color stresses continue indefinitely.
An invariant, coordinate-independent characterisation of the chaos is
provided by means of fractal basin boundaries.
\end{abstract}
\vskip 5truept
] 

\narrowtext
\begin{picture}(0,0)
\put(380,170){{ CfPA-97-TH-06}}
\end{picture} 
\setcounter{section}{1}

Yang-Mills fields are central to quantum theories of elementary particles.
They are of interest to dynamicists since they evolve chaotically in flat
spacetime \cite{{bms},{mss},{cs},{cs2}}. But how do Yang-Mills fields behave
in the early universe? Does the chaos persist, or is it eradicated by the
general relativistic effects of cosmological expansion? The earliest studies
of the Einstein-Yang-Mills (EYM) system assumed isotropic cosmological
expansion. This imposes special symmetries on the dynamics and the dynamical
system is integrable: chaos cannot exist \cite{gv}. Recently, a Yang-Mills
theory was formulated in an axisymmetric, spatially homogeneous universe 
\cite{dk}. The large number of degrees of freedom made an analysis of the
full dynamics difficult, and the coordinate dependence of the standard
chaotic indicators meant that the relativistic chaos could not be
invariantly characterised.

In this letter, we analyse general relativistic Yang-Mills chaos using
invariant topological methods. We make the problem more tractable by an
economical definition of variables, which reduces the dimension of the
phase space
substantially (from $8-D$ to $5-D$). In the physical picture that
emerges, the asymptotic evolution of the spatial volume of the universe
imitates a radiation-dominated universe, while the shear diminishes
chaotically.

Previous studies of chaos in general relativity have focused on the
non-axisymmetric Bianchi type VIII and IX (Mixmaster) universes, where the
presence of anisotropic 3-curvature creates an infinite sequence of chaotic
oscillations on approach to an initial Weyl curvature singularity at $t=0$ 
\cite{{mix},{kl2},{barrow}}. This behaviour is intrinsically general
relativistic. By contrast, the chaotic EYM cosmology that we study is
different: chaos exists even when the metric is axisymmetric and the
curvature is isotropic.

We evolve the Yang-Mills fields in the simplest anisotropic metrics of
Bianchi type I. The color degrees of freedom of the Yang-Mills gauge fields
oscillate chaotically, while the expansion attenuates their overall energy.
The EYM action is 
\begin{equation}
S=\int d^4x\sqrt{-g}\left[ -{\frac 1{16\pi G}}{\cal R}-{\frac 14}F_{\mu \nu
}F^{\mu \nu }\right] \ \ 
\end{equation}
where $F_{\mu \nu }$ is the gauge-invariant field strength. The spacetime is
described by the axisymmetric Bianchi I metric, with scale factors $b(t)$
and $c(t)$: 
\begin{equation}
ds^2=-dt^2+b^2(t)\left( dx^2+dy^2\right) +c^2(t)dz^2.
\end{equation}
After fixing the internal gauge of the Yang-Mills field, the matter can be
parametrized by two variables $(\alpha ,\gamma )$ \cite{dk}, which can be
thought of as color degrees of freedom for the massless gauge fields. The
cosmological evolution is an orbit in the $(b,c,\alpha ,\gamma )$ phase
space. 

We define the mean expansion scale factor by $a\equiv (b^2c)^{1/3},$ the
shear anisotropy by $\chi \equiv (b/c)^{1/3}$, with volume expansion $%
H_a\equiv \dot a/a,$ and shear $H_\chi \equiv \dot \chi /\chi $. The scaled
Yang-Mills field strengths will be defined by $(\Psi ,\Gamma )\equiv (\alpha
/(a\chi ),\gamma \chi ^2/a),$ with conjugate momenta $(\Pi _\Psi ,\Pi
_\Gamma )=(\dot \alpha /(a\chi ),\dot \gamma \chi ^2/a)$. The Einstein
equations reduce to 
\begin{eqnarray}
\dot H_a+{\ 2}H_a^2+H_\chi ^2 &=&0  \label{dha} \\
\dot H_\chi -H_\chi ^2+H_a^2+{3}H_aH_\chi &=&{\frac 12}\Psi ^4+{\frac 12}\Pi
_\Gamma ^2\ \ .  \label{dhchi}
\end{eqnarray}
The Hamiltonian constraint equation is 
\begin{equation}
{\frac 12}\Pi _\Gamma ^2+\Pi _\Psi ^2+{\frac 12}\Psi ^4+\Psi ^2\Gamma
^2=3\left( H_a^2-H_\chi ^2\right) \ \ .  \label{ham}
\end{equation}
The conservation of gravitational-plus-matter energy appears in the ($\Psi
,\Gamma $) subsystem as a loss of energy, given by $E(t)=3(H_a^2-H_\chi ^2)$%
. The matter conservation equations are 
\begin{eqnarray}
\dot \Pi _\Psi +2(H_a+H_\chi )\Pi _\Psi +\Psi \left( \Gamma ^2+\Psi
^2\right) &=&0  \nonumber \\
\dot \Pi _\Gamma +2(H_a+H_\chi )\Pi _\Gamma +2\Psi ^2\Gamma &=&0 \ .
\end{eqnarray}
The matter sector is thus a driven, dissipative system. The Yang-Mills
coordinates decay adiabatically due to the expansion, but they are also
driven by oscillations in $H_\chi $.

There are 4 coordinates, $(H_a,H_\chi ,\Psi ,\Gamma ),$ and 8 degrees of
freedom. This system permits at most $4$ positive Lyapunov exponents.
However, eqns (\ref{dha}) and (\ref{dhchi}) are constraints. Consequently,
only $6$ constants of integration need be specified. The Hamiltonian
constraint reduces this by one, leaving $5$ degrees of freedom. If each of
the three constraint equations are unique, then there is only one positive
Lyapunov exponent, as in flat spacetime ($H_a=H_\chi =0)$. This is to be
expected since the vacuum Bianchi I universe is integrable. Any chaos in the
metric variables is therefore a consequence of chaos in the Yang-Mills
field. This was also found to be the case in Ref. \cite{cos}.

The color amplitudes of the Yang-Mills system scatter around the potential
of Fig. \ref{pot}. The chaos is more complicated than that found in
Mixmaster universes because of the rapidly varying curvature of the
hyperbolic potential walls. In flat spacetime the energy remains constant,
and the contours define different surfaces of constant energy. In the
Bianchi I universe of general relativity, $E(t)$ drops asymptotically (see
eqn (\ref{ham})). The four hyperbolic potential walls shrink as the energy
available to the Yang-Mills field decays. We see from Fig. \ref{traj}, the
trajectories occupy smaller and smaller volumes of the matter phase-space.

Chaos is often quantified by computing Lyapunov exponents. However, the
values of Lyapunov exponents are coordinate dependent in general relativity 
\cite{{rugh}}, because of the coordinate covariance of Einstein's equations
(another manifestation of the so called `problem of time' in cosmology). The
authors of Ref. \cite{dk} cite this fact, together with the large number of
degrees of freedom (which lead to Arnold diffusion) and the non-compact
phase space, as major obstacles to a generalization of the study of chaotic
properties of Yang-Mills theory to curved spacetimes. These barriers can be
overcome by the methods of chaotic scattering \cite{{carl},{cos},{nj},{cg}},
which allow us to identify fractal sets which fully characterize the chaos.
Fractals cannot be hidden by a reshuffling of coordinates: their existence
and dimension are coordinate-independent topological features \cite
{{carl},{cos},{nj},{cg}}. The fractal set that we seek is the set of all
periodic orbits: by analogy with the accessible states in thermodynamics,
this periodic set completely describes the chaotic dynamics. The fractal set
is also called a `strange repellor', or `strange saddle'. If the global
expansion of the universe could be projected out of the dynamical evolution,
the repellor would consist of all periodic orbits trapped in the potential.
When the global expansion is included, the periodic orbits are actually
self-similar, reminiscent of the dynamics of the Bianchi IX cosmology \cite
{{nj},{mix},{kl2},{barrow},{berger},{burd}}.

The strange repellor can be coded by a symbolic dynamics, or isolated
directly by numerical techniques such as the PIM method. It is easier to
employ the method of fractal basin boundaries. 
A slice of initial conditions is taken through phase space. All possible
asymptotic states are determined and assigned a color (black or white here).
If large blocks of initial data space suffer the same fate, then the outcome
basin will look smooth and monochromatic; by contrast, highly mixed,
fractalized basins indicate a sensitivity to initial conditions, as well as
mixing and folding of trajectories. Hence, fractalized basins signal chaos
in a covariant way. All observers agree upon the occurrence of the events
used to construct the fractal, and all will agree on the dimension of the
basin boundary.

A typical trajectory will travel down one of the $\Gamma $ channels before
rebounding back into the scattering region of the potential. This is again
reminiscent of the Mixmaster system, where the repellor was also inefficient 
\cite{nj}. In Mixmaster, the repelling set can be isolated by artificially
opening the exit pockets, so that orbits thrown far from the scattering
region were allowed to escape. Similarly, here, we cut holes in the pockets
and assign the color white to the initial condition if the orbit falls down
the upper pocket ($\Gamma >0$), and black if it falls down the lower pocket (%
$\Gamma <0$). In flat spacetime, this is straightforward. In the Bianchi I
spacetime, the potential walls move inward as the energy is not constant.
However, the same procedure can be followed, so that the angular size of the
hole in the pocket is the same for all energy contours. Basin boundaries,
sliced through $(\Psi ,\Gamma )$ coordinates, are shown in Fig. \ref{bb}.
The pockets are of size $\theta =2\arctan 0.5$. The fractal basins look very
similar to those we find in a flat spacetime. The box-counting dimension was
estimated to be maximal, namely, $D_o=2$ (the same dimension as we found for
the basins in flat spacetime). If the pockets are made infinitesimally thin,
the fractal boundary fills all of phase space.

The interweaving of outcomes, demonstrated by 
the basin boundary structure, corresponds to a final-state sensitivity that
is directly related to the fractal dimension. An $\epsilon $ uncertainty in
initial values leads to a final-state uncertainty of $\epsilon ^{N-D_o},$
where $N$ is the dimension of the phase space. In the $N=2$ slice of Fig. 
\ref{bb}, we found $D_o=2$. Axisymmetric, Bianchi type I, EYM cosmologies
are therefore very sensitive to initial conditions, and the shear evolution
is highly chaotic.

The chaotic scattering has a subtle effect on the large-scale structure of
the spacetime. In vacuum, the scale factor grows as $a\propto t^{1/3},$ and
the shear is of comparable importance, $\chi \propto t^{1/3}$. The color
oscillations scatter the metric variables chaotically. As the universe
expands, the volume of the $(\Psi ,\Gamma )$ phase space redshifts as $\rho
\sim 1/a^4$. The scale factor eventually evolves towards the behaviour of a
radiation-dominated universe, $a\sim t^{1/2},$ and is unaffected by the
chaotic color oscillations. 
The relaxation timescale for this behaviour is related to the Lyapunov
exponent, and hence to the fractal dimension \cite{foot}.
A rough estimate indicates that trajectories attract onto $H_a=1/(2t)$ after
a few e-folds of the scale factor. The shear is more vulnerable to
stochastic behaviour and the color oscillations sustain oscillations of $%
H_\chi $ about zero. This general behaviour can be seen by examining the
scattering angle in minisuperspace. A trajectory leaves the $(\ln a,\ln \chi
)$ minisuperspace plane at an angle $\theta ={\rm arctan}(H_\chi /H_a)$.
Fig. \ref{scat} compares the scattering angle in the case of regular motion
to the angle in the chaotic case. 
The scattering angle is not constant in time as can be seen in Fig. \ref{ang}%
. While $H_a$ escapes the effects of the chaotic scattering, $H_\chi $
remains sensitive and continues to oscillate.

The dependence of the final scattered angle on the impact parameter is a
direct probe of the strange repellor. In a simple chaotic system, the
scattering angle can become fractal. No matter how small a difference in the
incident velocity, the difference in scattered angle is sizeable. The
scattering angle and the basin boundaries are different ways of viewing the
same phenomenon. Regardless, the scattering angle provides an important
perspective. When the motion is regular, the angle approaches $\pm \pi /4$
and so $H_\chi \sim \pm H_a$. The evolution of the shear is as important as
the global expansion of the spacetime volume, and initially the spacetime is
highly anisotropic. At late times, under the influence of chaos, the
scattering angle clusters about zero, and $|H_\chi |\ll H_a$, with the shear
frozen at a nearly constant value. Since the field equations were shown to
be invariant under a rescaling of the shear, we can choose that scale so
that $\ln \chi \rightarrow 0$ at freeze out. Physically, this is equivalent
to the statement that $a\sim b\sim c$. The anisotropy decays and the final
state looks isotropic.

The chaotic behaviour influences the solution only through the second-order
modes and the expansion of the volume behaves as if the universe is
radiation dominated. This reveals a further connection between the chaotic
behaviour in flat space-time and in expanding universes. If the
energy-momentum tensor is trace-free, as it is for Yang-Mills fields, then
any solution of the equations of motion in flat spacetime can be conformally
rescaled to obtain a solution of the equations of motion in an isotropically
expanding universe. This solution is approximate since it neglects the
back-reaction of the motions on the space-time metric. Though we did not
actually employ this approximate method, such a scaled solution 
would be an increasingly good description of the late-time behaviour, since
isotropization occurs in that limit.

The global scale factor 
is fairly impervious to the buffeting of the other degrees of freedom. Small
shear and color oscillations will continue forever, getting ever smaller in
amplitude, and only asymptotically diluting to zero. We expect an infinite
number of oscillations to occur to the future of any finite time. However on
the coarsest scales, the universe evolves as though filled with radiation.
Most interestingly, the original anisotropy is eroded and the universe
appears isotropic.

\vskip 5truept

JJL is especially grateful to N. Cornish and P. Ferreira for valuable
discussions. JDB is supported by the PPARC and acknowledges support from the
Center for Particle Astrophysics, Berkeley. JJL is supported in part by a
President's Postdoctoral Fellowship and acknowledges support from the PPARC
at the Astronomy Centre, Sussex.

\vfill\eject

{

\widetext

\
\begin{figure}[h]
\vspace{56mm}
\includegraphics{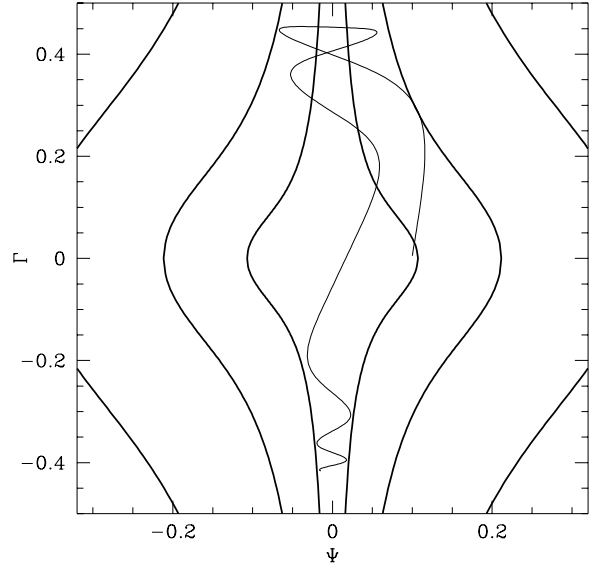}
\vspace{15mm}
\caption{
Isocontours of the color potential (eqn. (1.6)). As the energy, 
$E(t),$ decreases, the walls of the potential move inward and the color
oscillations are confined accordingly. A typical trajectory is shown.
\label{pot}}  \end{figure} 

\vskip 50truept

\
\begin{figure}[h]
\vspace{56mm}
\includegraphics{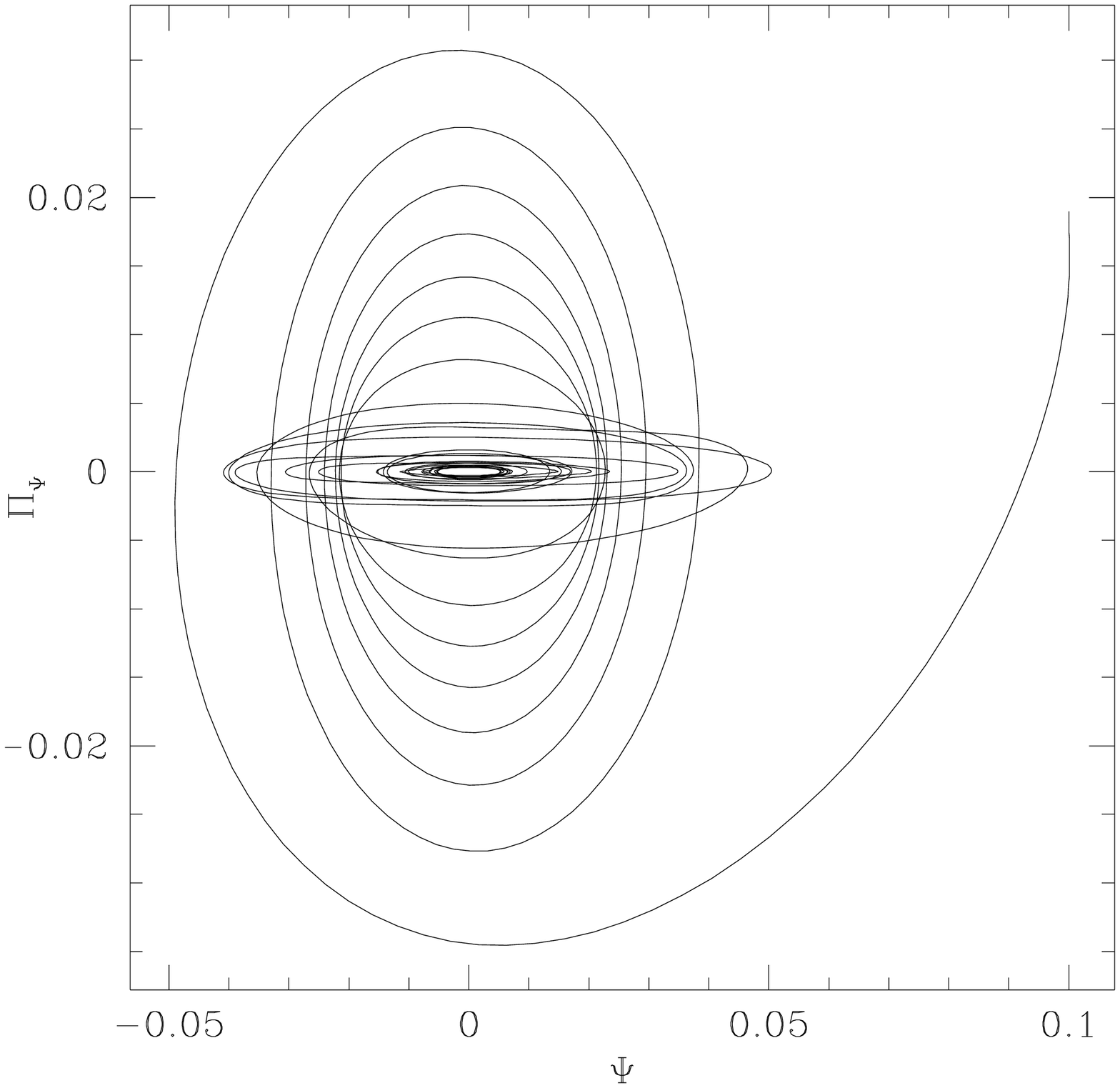}
\includegraphics{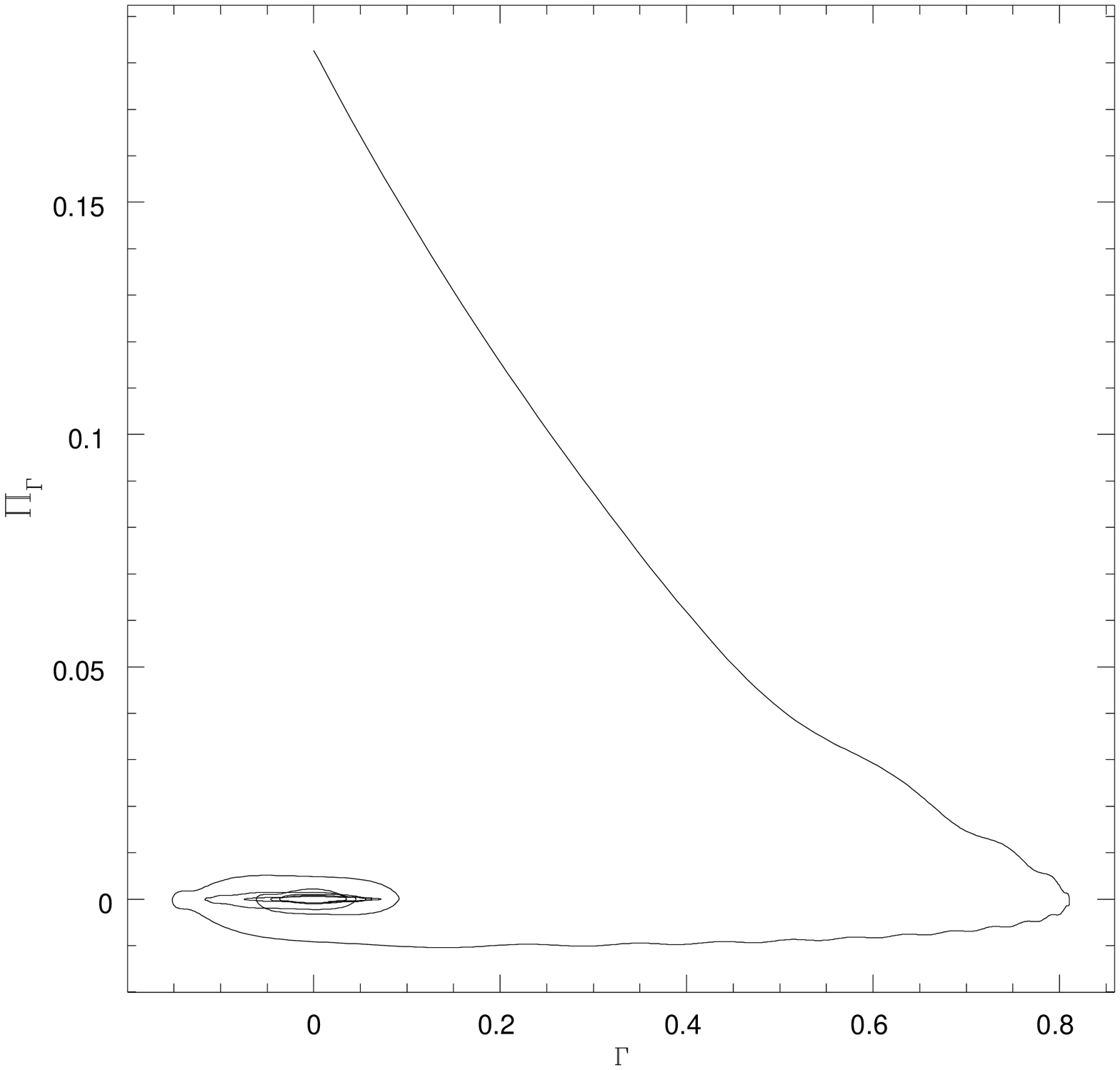}
\vspace{15mm}
\caption{A projection of the Bianchi I phase space onto the $(\Psi ,\Pi _\Psi
)$ plane (left) and onto the $(\Gamma ,\Pi _\Gamma $) plane (right). The
initial values for this trajectory are $\Psi =0.1,\Gamma =0,H_a=0.11,H_\chi
=0.08,\dot \Psi =0$ and $\dot \Gamma $ is fixed by the Hamiltonian
constraint, eq. (1.5).
\label{traj}} \end{figure}

\vfill\eject
\subsection{}
\vfill\eject

\
\begin{figure}[h]
\vspace{56mm}
\includegraphics{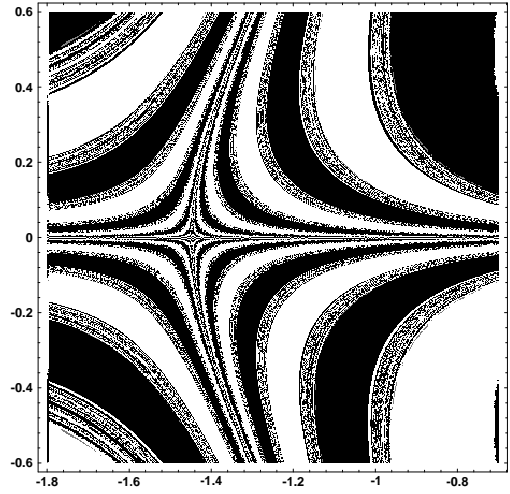}
\vspace{15mm}
\caption{The strange repellor is revealed in these fractal
basin boundaries.  Holes are cut in the $\Gamma $ channels 
to allow typical trajectories to escape.  The angular size of the
pockets in this simulation is $\tan(\theta/2)=0.5$.\label{bb}}  \end{figure} 

\vskip 50truept

\
\begin{figure}[h]
\vspace{50mm}
\includegraphics{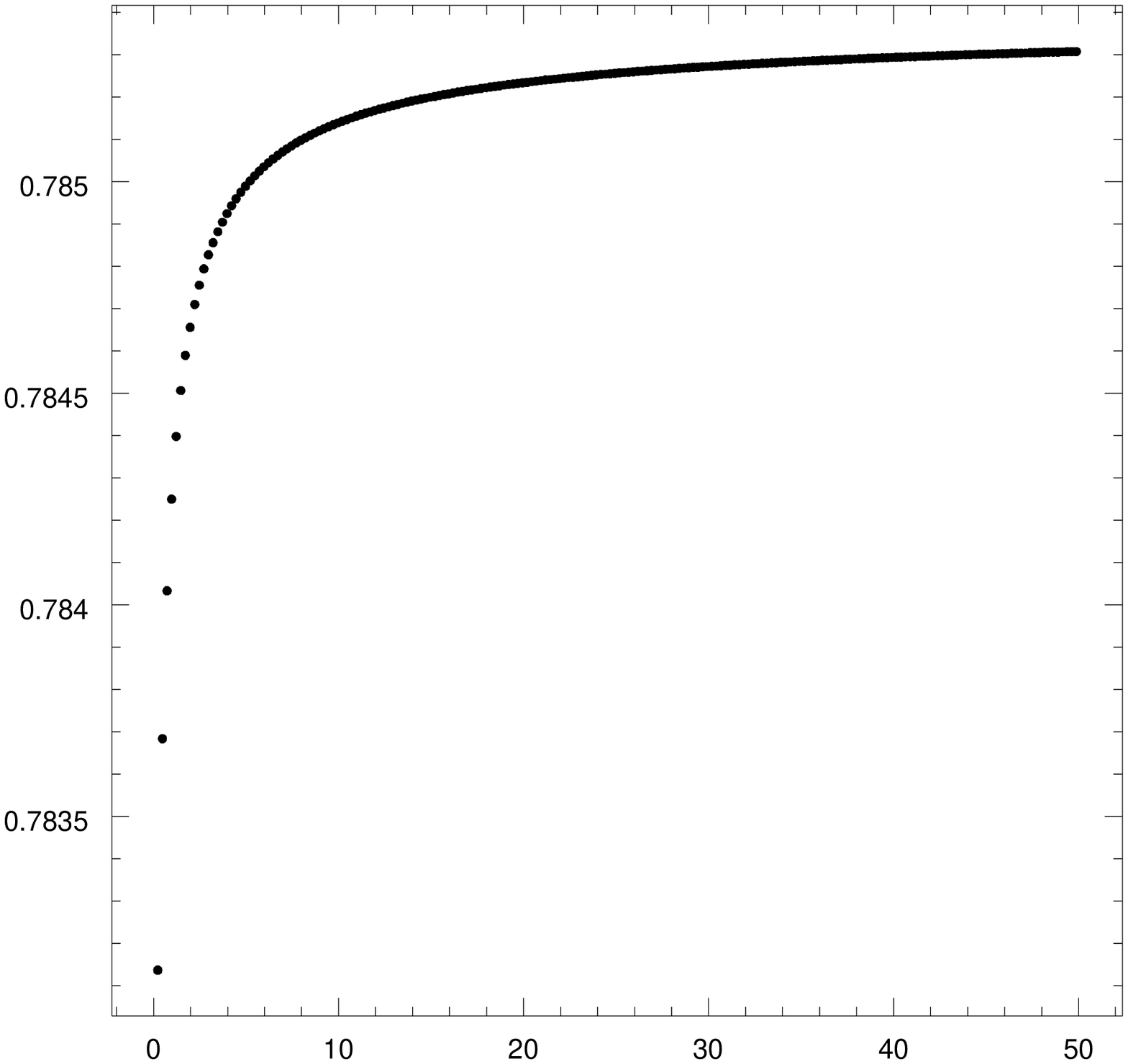}
\includegraphics{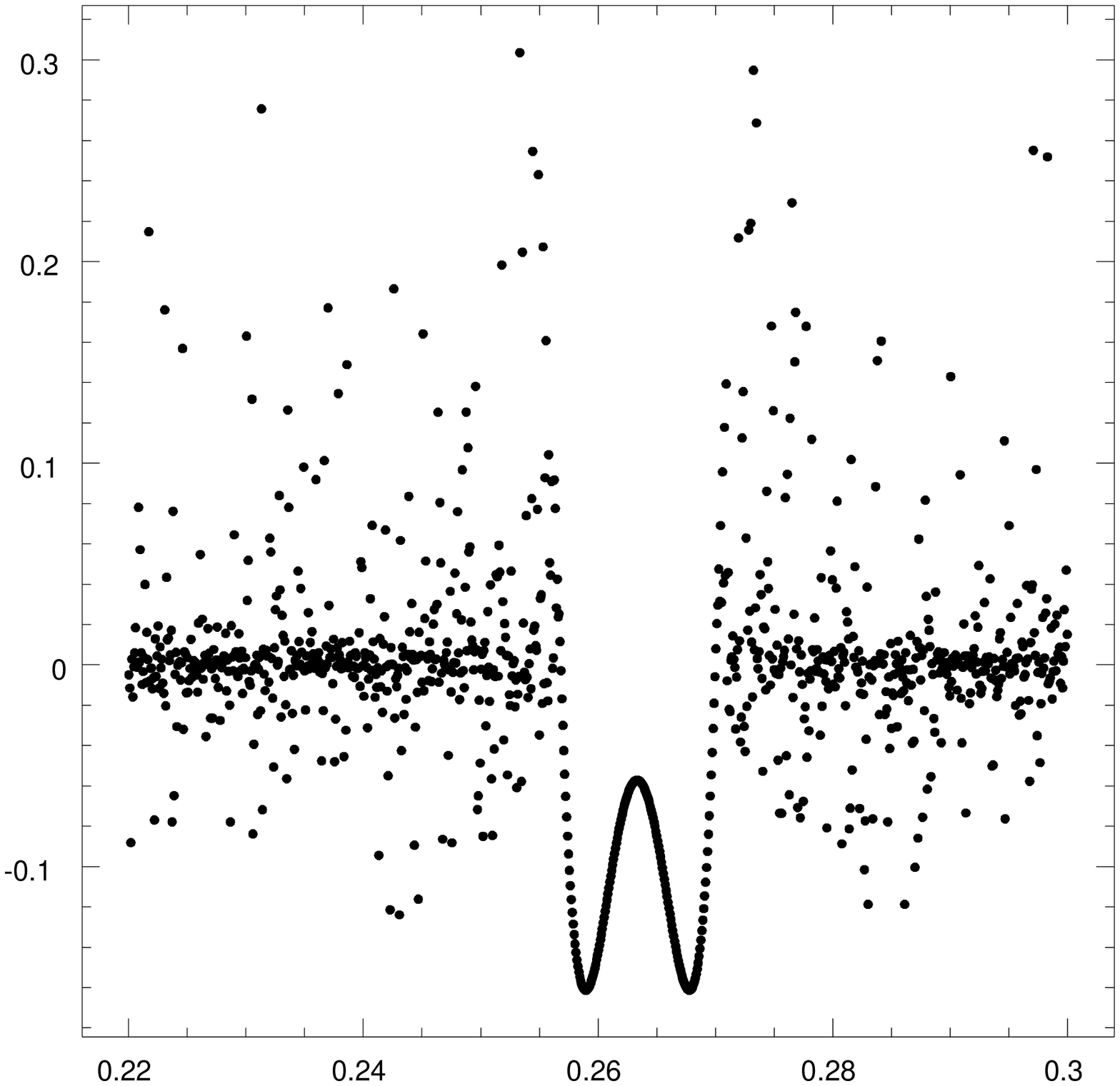}
\vspace{25mm}
\caption{
The scattering angle defined as $\theta={\rm arcsin}(H_\chi/H_a)$
as a function of the initial $H_{a}$.
The left plot is not chaotic with $\Psi=\dot \Psi=0$.
In the right panel
the initial
conditions for the remaining degrees of freedom
are $H_{\chi o}=0.0999$,$\Psi=0.1=-\Gamma=-\dot \Psi$.
Again, the Hamiltonian constraint fixes $\dot \Gamma$.
\label{scat}}
\end{figure}

\vfill\eject
\subsection{}
\vfill\eject

\
\begin{figure}[h]
\vspace{56mm}
\includegraphics{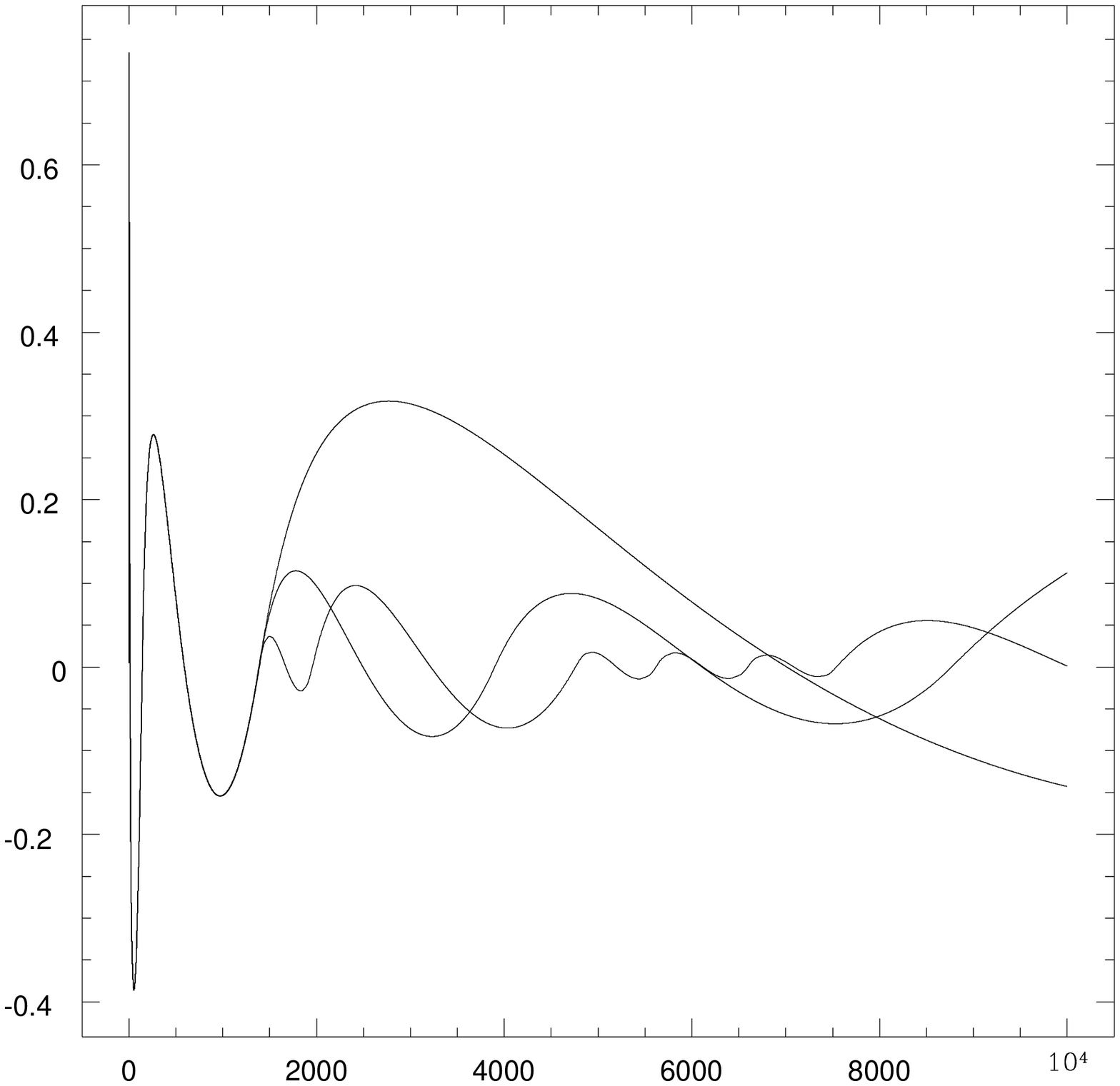}
\includegraphics{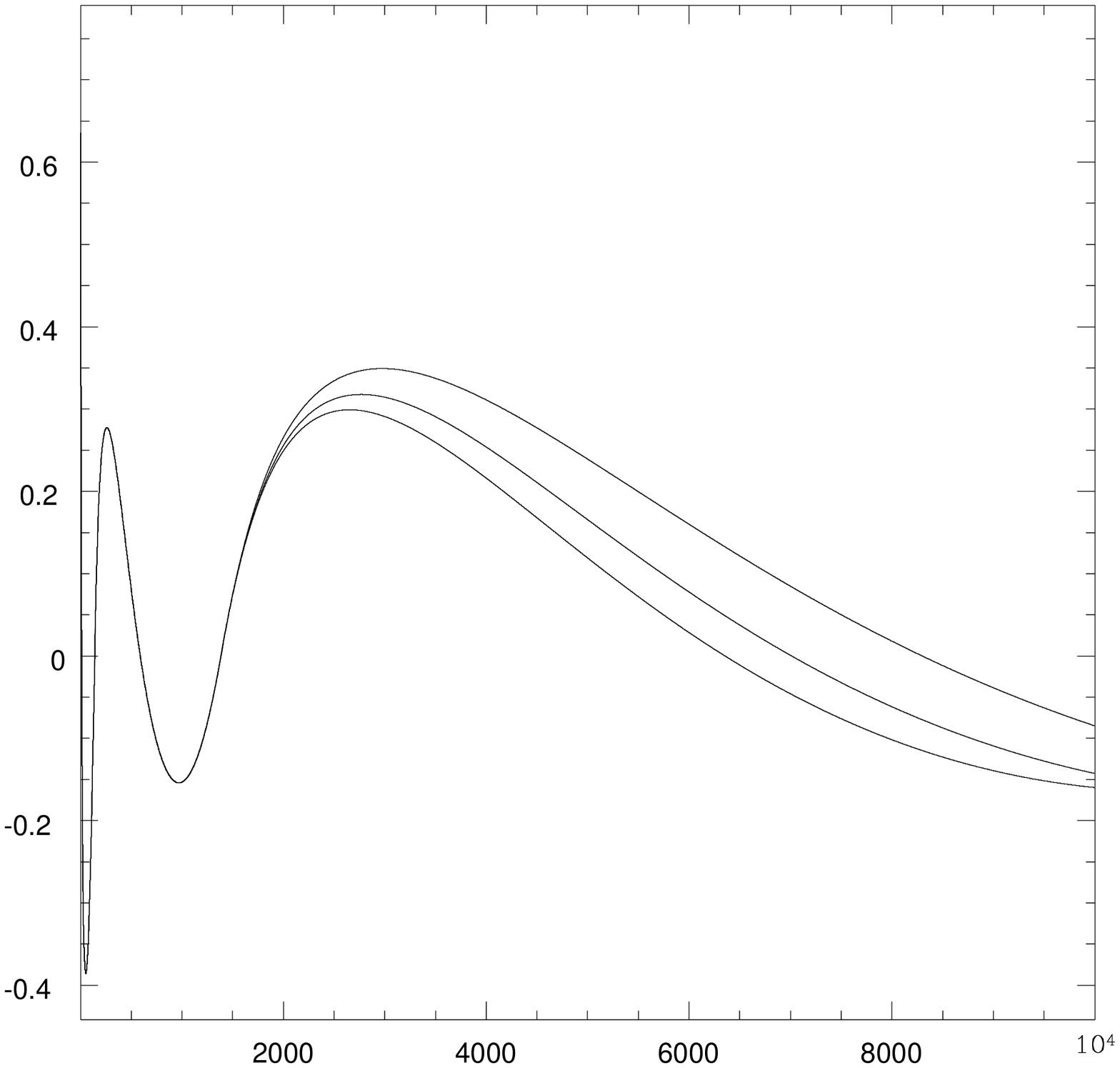}
\vspace{15mm}
\caption{The scattering angle 
as a function of cosmic time.
The initial
conditions 
are $H_{a}=20.0002,20.000245,20.00026$,
$H_{\chi }=0.0999$,$\Psi=0.1=-\Gamma=-\dot \Psi$.
The scattering angle 
as a function of cosmic time corresponding to the regular dip seen
in the above blow up in Fig. 4.
On the right, the initial conditions 
are $H_{a}=20.00026$,$20.000265$,$20.0002675$,
$H_{\chi }=0.0999$,$\Psi=0.1=-\Gamma=-\dot \Psi$.
The Hamiltonian constraint fixes $\dot \Gamma$. \label{ang}
}
\end{figure}

}


\begin{references}
\bibitem{foot}
Since the fractal dimension codifies the loss of information it is sensible
that the Lyapunov exponent determined in any gauge must be intimately
connected with this dimension. In fact it has been conjectured that the
information dimension of the repellor is directly related to the Lyapuov
dimension, a gauge invariant combination of the Lyapunov exponent and the
time scale of the chaotic transient \cite{{ky},{goy},{young},{kan}}. For a
demonstration of the equivalence in relativity see Ref. \cite{nj}.
\end{references}
\end{document}